\documentclass[reprint,aps,ams,amsmath,prx,twocolumn,longbibliography]{revtex4-1}
\usepackage{graphics}
\usepackage{graphicx}
\usepackage{epsfig}
\usepackage{amsmath}
\usepackage{amssymb}
\usepackage{amsfonts}
\usepackage{bm}
\usepackage{color}
\usepackage{bbm}
\usepackage{hyperref}
\usepackage[normalem]{ulem}
\usepackage{comment}
\usepackage{soul}
\usepackage{cleveref}

\begin{document}

\title{Near-field heat transfer and drag resistance in bilayers of composite fermions}

\author{Dmitry Zverevich}
\affiliation{Department of Physics, University of Wisconsin-Madison, Madison, Wisconsin 53706, USA}

\author{Alex Levchenko}
\affiliation{Department of Physics, University of Wisconsin-Madison, Madison, Wisconsin 53706, USA}

\date{January 29, 2024}

\begin{abstract}
Heat transfer is studied in the system of electron double layers of correlated composite fermion quantum liquids. In the near-field regime, the primary mechanism governing interlayer energy transfer is mediated by the Coulomb interaction of thermally-driven charge density fluctuations. The corresponding interlayer thermal conductance is computed across various limiting cases of the composite fermion Chern-Simons gauge theory, encompassing ballistic, diffusive, and hydrodynamic regimes. Plasmon enhancement of the heat transfer is  discussed. The relationship between the heat transfer conductance and the drag resistance is presented for electron states formed in the fractional quantum Hall effect of even denominator filling fractions.  
\end{abstract}

\maketitle

\section{Introduction}

In studies of strongly correlated electrons of condensed matter systems, the problem of two-dimensional (2D) fermions coupled to a gauge field arises naturally in many different contexts as an effective low-energy model. The emergent gauge theory provides a powerful framework to address properties of quantum spin systems, fractional quantum Hall liquids, and unconventional superconductors \cite{Nagaosa,Fradkin}. 

In the paradigmatic example of the fractional quantum Hall effect with the even denominator filling fractions, such as the half-filled Landau level \cite{KZ,HLR}, the effective action description is obtained in connection to the composite fermion (CF) representation \cite{Heinonen,Jain,Son:2015}. A composite fermion is introduced in the theory by attaching an even number of flux quanta to an electron. This transformation is realized by introducing an appropriate Chern-Simons gauge field. There are several advantages of such reformulation of the original problem. At the mean-field level, the composite fermion experiences no magnetic field, which is enforced by the cancellation of the properly chosen Chern-Simon flux and the externally applied magnetic field. Consequently, the system of strongly interacting electrons in a magnetic field can be mapped onto a Fermi liquid of weakly interacting composite fermion quasiparticles. Extending beyond the mean-field level, fluctuations of the gauge field can be addressed within the random-phase approximation (RPA). The approach based in the Chern-Simons theory has demonstrated success, enabling relatively straightforward and systematic calculations of experimentally measurable quantities such as conductivities and various response functions, as elaborated in Refs. \cite{Simon-PRB93,Kim-PRB94,Khveshchenko,Mirlin-PRL97,Mirlin-PRL98,Ludwig-PRB08}. It is worth noting, however, that the electromagnetic and thermal responses of composite fermions differ significantly from those observed in conventional Fermi liquids.

The double-layer quantum well heterostructures, composed of closely spaced parallel two-dimensional electron systems, unveil a myriad of intriguing quantum Hall physics. These systems offer a unique platform for exploring nonlocal transport effects that are exclusive to bilayers. Experimentally observed examples of emergent phenomena in high-mobility semiconductor devices or graphene bilayers include Coulomb drag \cite{Lilly-PRL98,Gramila-PRL00,Klitzing-PE00,Tutuc-PRB09,Price-PRB10} and quantum Hall drag \cite{Kellogg-PRL02,Tutuc-PRL04,Liu-PRL17,Liu-NP2017}, particularly at half-filling per layer. Additionally, superfluid exciton condensation of composite quasiparticles has been observed \cite{Nandi-NP2012,Eisenstein,Li-NP2017,Li-NP2019}. The existing theories provide a foundational support for many observed features \cite{Ussishkin-PRB97,Sakhi-PRB97,Ussishkin-PRL98,Millis-PE99,Bonsager-PRB00,Nayak-PRB01,Narozhny-PRL01,Stern-PRL02,Patel-PRB17}, and go beyond by identifying a host of other possible states at fractional total filling some of which are expected to display exotic topological properties \cite{Scarola,Alicea,Barkeshli,Zaletel,EA-Kim}. 

This paper is motivated by the physics of bilayers, with a primary objective of investigating specific thermal transfer properties defined by the near-field effect \cite{Volokitin,BenAbdallah,Bimonte}. It has long been established \cite{Rytov,PolderVanHove,Pendry} that the heat flux between two closely spaced planar bodies, maintained at different temperatures, is dominated by fluctuation-driven near-field evanescent electromagnetic modes. This regime is realized when the interlayer separation becomes smaller than the thermal de Broglie wavelength of the photon. Notably, for conducting layers, such a near-field effect is dominated by the Coulomb interactions between thermal fluctuations of electron density \cite{Mahan}. Despite its relevance to most modern nanostructures, see e.g. Refs. \cite{Kim-NFHT,Song-NFHT,St-Gelais-NFHT,Yang-NFHT}, this physics has yet to be explored in the context of composite fermion double-layer systems, constituting our primary goal.

The presentation is organized as follows. In Section \ref{sec:NFHT}, we apply the RPA of Chern-Simons theory to compute the near-field thermal conductance between unequilibrated layers in a composite fermion picture, discussing the coupling between layers in purely electronic terms. This approach elucidates the underlying physical picture and establishes connections to previous works on heat transfer between 2D conductors. In Section \ref{sec:Dis}, we employ the semiclassical Boltzmann equation to incorporate impurity scattering and derive heat transfer in the diffusive limit. Section \ref{sec:Hydro} delves into the hydrodynamic limit of the composite fermion liquid, assuming fast intralayer equilibration and subjecting the system to a disorder potential with a long correlation radius. In this regime, we derive heat conductance in terms of dissipative coefficients of the fluid, such as intrinsic conductivity and viscosity, and the correlation function of the disorder potential. Additionally, we highlight the mechanism of plasmonic enhancement. For a broader perspective, in Section \ref{sec:Drag} we draw parallels between the technicalities underlying the near-field effect and the problem of nonlocal Coulomb drag resistance \cite{CD-Review}. Finally, in Sec. \ref{sec:estimates} we provide typical parameters used in the calculations and the order of magnitude estimates. In the paper we work in the natural units setting Boltzmann and Planck constants to unity $k_B=\hbar=1$. 

\begin{figure}[t!]
\centering
\includegraphics[width=3in]{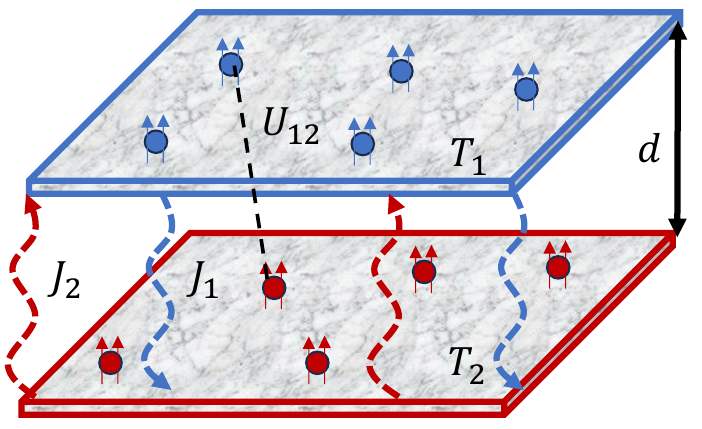}
\caption{Schematic representation of the interactively coupled two-dimensional composite fermion liquids formed in an electron double layer system with interlayer spacing $d$. 
The composite fermion quasiparticles are depicted as circles with attached fluxes $\phi$ labeled by arrows. The interlayer coupling is denoted by the Coulomb 
potential $U_{12}$. The underlying background of each layer represents disorder potential.   
Each layer is kept at different temperatures $T_{1,2}$ and wavy dashed lines pointing up or down represent vertical heat fluxes $J_{1,2}$ between the layers generated in the near-field regime.}
\label{fig:EDL-CF}
\end{figure}


\section{Near-field effect in bilayers}\label{sec:NFHT}

\subsection{Formalism}

The basic setup for considering the near-field effect consists of two parallel 2D layers separated by a distance $d$, and kept at different temperatures
$T_1$ and $T_2$, see Fig. \ref{fig:EDL-CF} for illustration. It can be shown that in such a bilayer the heat current per unit area is given by 
\begin{equation}\label{eq:J}
J=\int \omega[N_1-N_2]\Im\Pi_1(q,\omega)\Im\Pi_2(q,\omega)|U_{12}(q,\omega)|^2d\Gamma_{q\omega}
\end{equation}
In the theory of near-field heat transfer (NFHT) this expression is often called the Caroli formula. It can be derived from variety of methods based on e.g. fluctuational electrodynamics \cite{Rytov,PolderVanHove}, nonequilibrium Green’s function formalism \cite{JiangWang-PRB17,ZhangWang-PRB18}, kinetic equation \cite{Ying-PRB20,Hekking-PRB20,Kamenev-LTP}, and Ehrenfest theorem \cite{AL-PRB22,AL-NFHT}. The physical picture behind Eq. \eqref{eq:J} was comprehensively discussed in the literature by many authors. Here we briefly repeat main points to have a self-contained presentation. In the Coulomb limit NHFT arises from scattering of electrons in different layers resulting in finite momentum $	q$ and energy $\omega$ transfer. Thus the integration expands over the phase space $d\Gamma_{q\omega}=d\omega d^2q/(2\pi)^3$ and $N_{1,2}(\omega)=1/(e^{\omega/T_{1,2}}-1)$ denotes the Planck distribution function. In the linear response, the layer polarizability $\Pi_{1,2}(q,\omega)$ is given by the density-density correlation function, which relates the induced charge density to the total electric potential \cite{Mahan-Book}. The response function $\Pi_{1,2}(q,\omega)$ also determines the dynamically screened Coulomb potential both for intralayer and interlayer interaction. It follows from the matrix Dyson equation. Within the limits of RPA and written in the layer basis it reads 
\begin{equation}
\hat{U}=\hat{V}_q\circ\left(1+\hat{\Pi}\circ\hat{V}_q\right)^{-1}
\end{equation}         
where $\circ$ denotes matrix multiplication and 
\begin{equation}
\hat{V}_q=V_q\left(\begin{array}{cc}1 & e^{-qd} \\ e^{-qd} & 1\end{array}\right), \quad 
\hat{\Pi}=\left(\begin{array}{cc} \Pi_1 & 0 \\ 0 & \Pi_2\end{array}\right).
\end{equation}
Here $V_q=2\pi e^2/\epsilon q$ is the bare Coulomb potential and $\epsilon$ is the dielectric constant of the host material surrounding
the electron layers. The fact that the polarization operator has no off-diagonal elements, reflects the assumed absence of tunneling between the layers.
In Eq. \eqref{eq:J} $U_{12}(q,\omega)$ is the off-diagonal element of $\hat{U}(q,\omega)$. It takes the form 
\begin{equation}\label{eq:U}
U_{12}=\frac{V_q e^{-qd}}{(1+V_q\Pi_1)(1+V_q\Pi_2)-V^2_q\Pi_1\Pi_2e^{-2qd}}.
\end{equation}

The validity of Eq. \eqref{eq:J} is not limited to a small difference between $T_1$ and $T_2$, however it is useful to introduce the linear in $\Delta T=T_1-T_2$ heat transfer conductance 
\begin{equation}
\varkappa=\lim_{T_{1,2}\to T}\frac{J(T_1,T_2)}{T_1-T_2}
\end{equation}
Differentiating Eq. \eqref{eq:J}, and assuming for simplicity identical layers, $\Pi_{1,2}=\Pi$, one finds 
\begin{equation}\label{eq:kappa}
\varkappa=\int \frac{\omega^2(\Im\Pi(q,\omega))^2|U(q,\omega)|^2}{4T^2\sinh^{2}(\omega/2T)}d\Gamma_{q\omega}. 
\end{equation}

In general, the temperature dependence of $\varkappa$ comes from the thermal factor and temperature dependence of the polarization function. For example, in a two-dimensional electron gas (2DEG) at temperatures below Fermi energy, polarization is weakly temperature dependent, therefore, the main dependence on $T$ comes from the thermal broadening and phase space available to quasiparticle excitations. In contrast, in graphene devices close to charge neutrality, polarizability is very strongly temperature dependent, which leads to additional features. The dependence of $\varkappa$ on the interlayer spacing $d$ is primarily determined by the screening effects, which are implicit in the form of the interlayer interaction potential $U(q,\omega)$.     

In bilayers of composite fermions, polarization is approximately temperature independent. However, properties of this function at low-frequency and long-wavelength lead to an emergence of the $T$-dependent momentum scale that sets the typical momentum transfer between the layers. In a conventional 2DEG this scale is simply set by the interlayer separation $q\sim 1/d$. Since $\Pi(q,\omega)$ also enters the interlayer interaction and modifies screening, this combined effect leads to a unique $T$ and $d$ dependence of $\varkappa$, which differs significantly from that of the Fermi liquid regime.  

\subsection{Composite fermion description}

The single-layer electronic polarizability $\Pi(q,\omega)$ can be calculated with the composite fermion theory. Following Ref. \cite{HLR} we introduce electronic $\hat{\Pi}^{\text{el}}$ and composite fermion $\hat{\Pi}^{\text{cf}}$ density and current response functions. The latter describes the response of the composite fermions to the total scalar and vector potentials including external, Coulomb, and Chern-Simons contributions.  These functions are related to each other,     
\begin{equation}\label{eq:Pi-Pi}
(\hat{\Pi}^{\text{el}})^{-1}=\hat{C}+(\hat{\Pi}^{\text{cf}})^{-1}, 
\end{equation}
by the Chern-Simons interaction matrix 
\begin{equation}
\hat{C}=\left(\begin{array}{cc} 0 &  2\pi i\phi/q \\ -2\pi i\phi/q & 0
\end{array}\right).
\end{equation}
For the case of the half-filled Landau level $\phi=2$. The composite fermion matrix $\hat{\Pi}^{\text{cf}}$ can be calculated within the RPA and simply corresponds to the noninteracting fermion at zero magnetic field. Therefore, it is a diagonal matrix
\begin{equation}
\hat{\Pi}^{\text{cf}}=\left(\begin{array}{cc} \Pi^{\text{cf}}_{\rho\rho} & 0 \\ 0 & \Pi^{\text{cf}}_{jj}
\end{array}\right).
\end{equation}
The density element of this matrix reflects the finite compressibility of the system. The current element describes the diamagnetism. In the low-energy limit $q/k_{\text{F}}\ll1$ and $\omega/v_{\text{F}}q\ll1$, and to the lowest nonvanishing order  \cite{HLR}
\begin{equation}\label{eq:Pi-rho-j}
\Pi^{\text{cf}}_{\rho\rho}\approx\frac{m^*}{2\pi} ,\quad \Pi^{\text{cf}}_{jj}\approx-\frac{q^2}{24\pi m^*}+\frac{i\omega k_{\text{F}}}{2\pi q},
\end{equation}
where $m^*$ is the quasiparticle effective mass \cite{Shayegan,Gervais},  $v_{\text{F}}$ and $k_{\text{F}}=\sqrt{4\pi n}$ are the Fermi velocity and momentum, respectively, with $n$ being average particle density per layer. The imaginary part of $\Pi^{\text{cf}}_{jj}$ describes Landau damping. We need to extract the density element of $\hat{\Pi}^{\text{el}}$ that enters Eq. \eqref{eq:kappa}. Inverting matrix in Eq.\eqref {eq:Pi-Pi} one finds 
\begin{equation}\label{eq:Pi-el}
\Pi(q,\omega)\equiv\Pi^{\text{el}}_{\rho\rho}=\frac{\Pi^{\text{cf}}_{\rho\rho}}{1-\Pi^{\text{cf}}_{\rho\rho}\Pi^{\text{cf}}_{jj}(2\pi\phi/q)^2}.
\end{equation}
Using the explicit forms of $\Pi^{\text{cf}}_{\rho\rho}$ and $\Pi^{\text{cf}}_{jj}$ from Eq. \eqref{eq:Pi-rho-j} we arrive at the well-known result \cite{HLR}
\begin{equation}\label{eq:Pi}
\Pi(q,\omega)=\frac{\nu q^3}{q^3-2\pi i\nu\phi^2\omega k_{\text{F}}},
\end{equation}
where we introduced the thermodynamic compressibility $\nu=(m^*/2\pi)/(1+\phi^2/12)$. The above form of the polarizability is reminiscent of that of a disordered electron gas $\nu Dq^2/(Dq^2-i\omega)$, see further discussion in Sec. \ref{sec:Dis}, but with an effective diffusion constant $D$ that depends linearly on $q$. This particular feature leads to a slow spreading of charge fluctuations and it also weakens screening effects thus enhancing the thermal transfer due to Coulomb coupling.     

\begin{figure}[t!]
\centering
\includegraphics[width=3in]{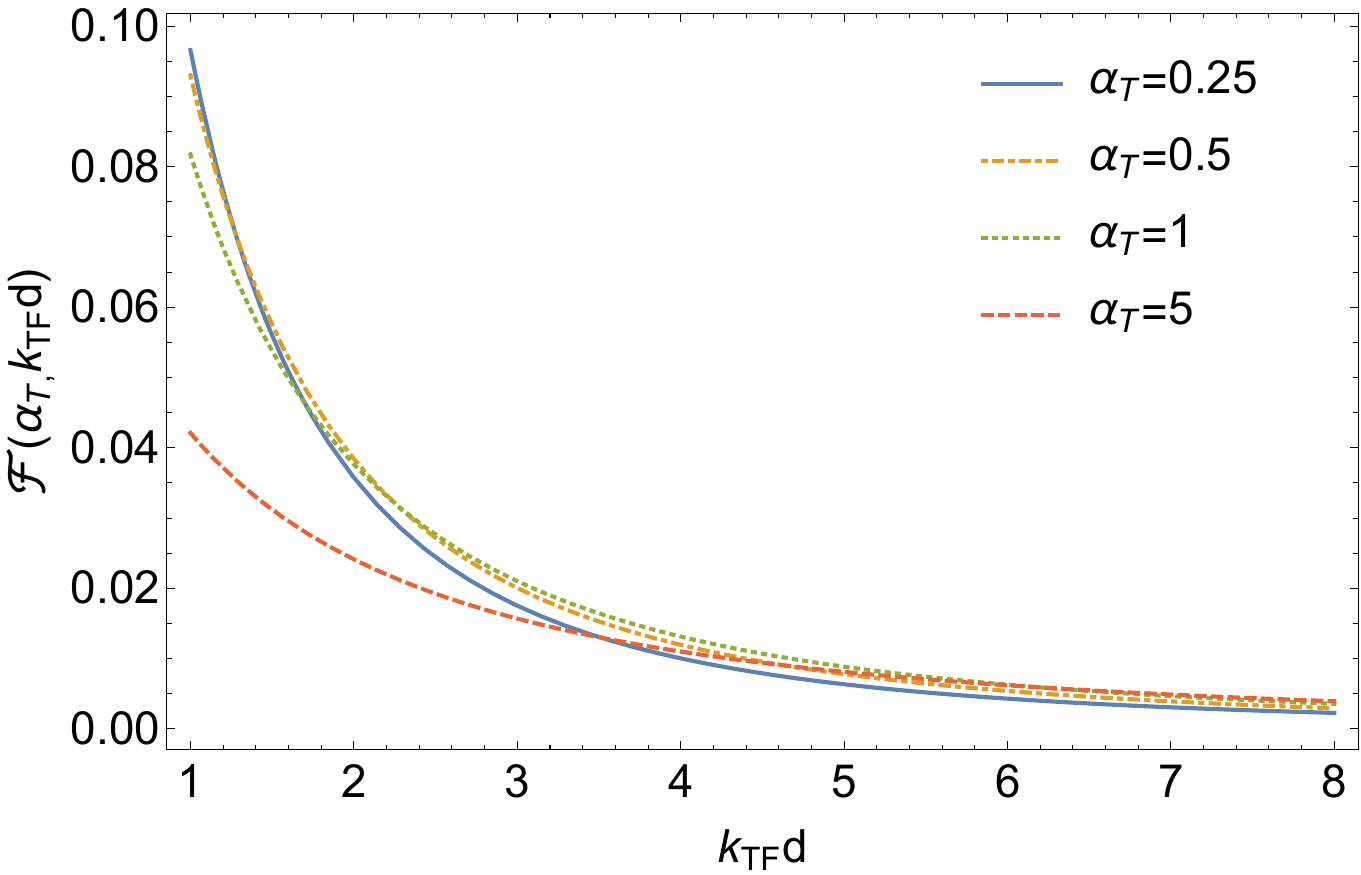}
\includegraphics[width=3in]{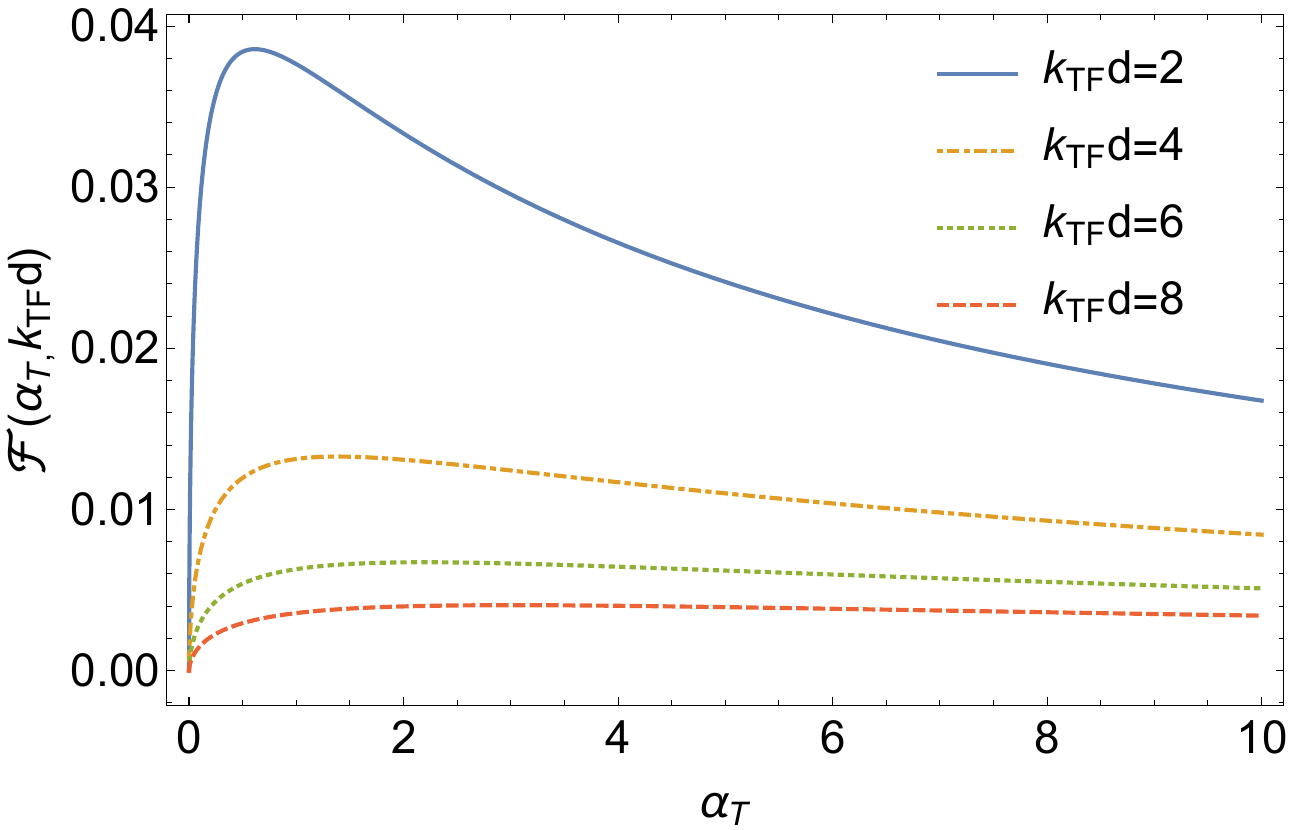}
\caption{Plots of the dimensionless function $\mathcal{F}$ defined by Eq. \eqref{eq:F} that show its dependence on the interlayer separation $d$ at different temperatures as described by the dimensionless parameter $\alpha_T$ (top panel) and vise versa, dependence on $\alpha_T$ at several values of $k_{\text{TF}}d$ (bottom panel). }
\label{fig:F-cf}
\end{figure}

\subsection{NFHT conductance}

With all the ingredients described in the previous section, we focus our attention to Eq. \eqref{eq:kappa}. The key element that we need to unpack is the product $\Im\Pi(q,\omega) |U(q,\omega)|$. In the case of symmetric layers, the denominator of the interlayer interaction in Eg. \eqref{eq:U} factorizes into a product of two simple terms $(1+V_q\Pi)\pm V_q\Pi e^{-qd}=1+V_q\Pi(1\pm e^{-qd})$. This leads to the following identity (omitting $q,\omega$ arguments of $\Pi$ and $U$ for brevity)
\begin{align}\label{eq:Pi-U}
&\Im(\Pi) |U|\notag= \\ 
&-\Im (\Pi^{-1})\left|\frac{V_qe^{-qd}}{[\Pi^{-1}+V_q(1+e^{-qd})][\Pi^{-1}+V_q(1-e^{-qd})]}\right|.
\end{align}

It is instructive to examine poles of the product $\Im(\Pi) |U|\notag$. From the expression above it is clear that they occur when $\Pi^{-1}=-V_q(1\pm e^{-qd})$. From Eq. \eqref{eq:Pi} it then follows that in the long wave length limit, $q\to0$, we have two overdapmed modes, $\omega_+\propto iq^2$ and $\omega_-\propto iq^3$. Therefore, density fluctuations relax slowly. It should be also stressed, that screening of such fluctuations is weak. These two factors lead to a substantial enhancement of heat transfer and drag resistance.  Based on these considerations one should expect an unconventional temperature dependence of $\varkappa$. Indeed, the typical frequency of excitations is set by temperature $\omega\propto T$. This immediately leads to a characteristic transfer momentum $q\propto T^{1/3}$ of the $\omega_-$ mode, which will be shown to dominate the response of the system. This is distinct from the case of conventional 2DEG, where typical $q$ are determined by the interlayer spacing $q\propto 1/d$. As a consequence, the phase space integral in Eq. \eqref{eq:kappa} will be dominated by the characteristic momenta of the collective modes. Since $q\propto T^{1/3}$ one expects $\varkappa$ to scale with the fractional power of temperature. The scaling $q\propto T^{1/3}$ at lowest temperatures also supports the argument of weak screening since in that limit $e^{-qd}$ factor in the interlayer interaction can be approximated by unity since $qd\ll1$.         

To further illuminate the physical significance of poles, one can express the polarization function $\Pi(q,\omega)$ in terms of the conductivity $\sigma(q,\omega)$. This is easily done by using the continuity equation relating density and current, along with the Kubo formula that defines conductivity from the current-current correlation function, whereas $\Pi$ is determned by the density-density correlation function. One thus finds $\sigma(q,\omega)=-ie^2(\omega/q^2)\Pi(q,\omega)$. Therefore, the dispersion relation for the collective modes can be written as 
\begin{equation}\label{eq:modes}
i\omega=\frac{q^2}{e^2}\sigma(q,\omega)V_q(1\pm e^{-qd}). 
\end{equation} 
In the range of frequencies where conductivity is real the solution to this equation gives only overdamped modes discussed above. However, when $\sigma$ is purely imaginary, the solution of Eq.\eqref{eq:modes} are given by plasmons of the double-layer system \cite{Fetter}, namely in-phase (optical) and out-of-phase (acoustic) density oscillations. Plasmons are higher in energy excitations and their role in the heat transfer and drag resistance will be discussed in Sec. \ref{sec:Hydro} and \ref{sec:Drag} respectively. We also remind that Eq. \eqref{eq:modes} played a pivotal role in the analysis of surface acoustic wave experiments \cite{Willett-PRL,Willett-PRB}, which provided the first clear indication of the existence of a compressible state of the half filled Landau level.       

\begin{figure}[t!]
\centering
\includegraphics[width=3in]{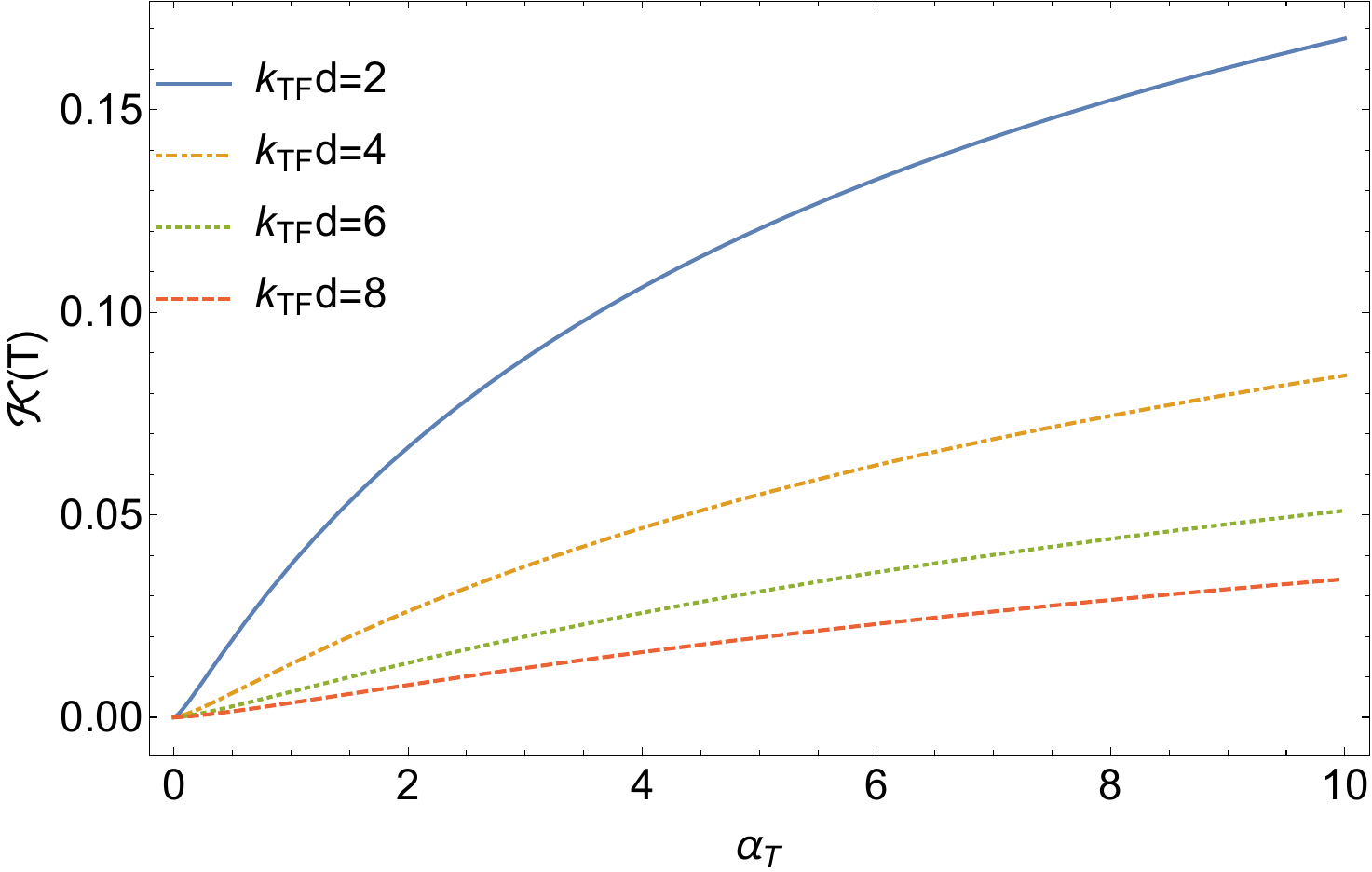}
\includegraphics[width=3in]{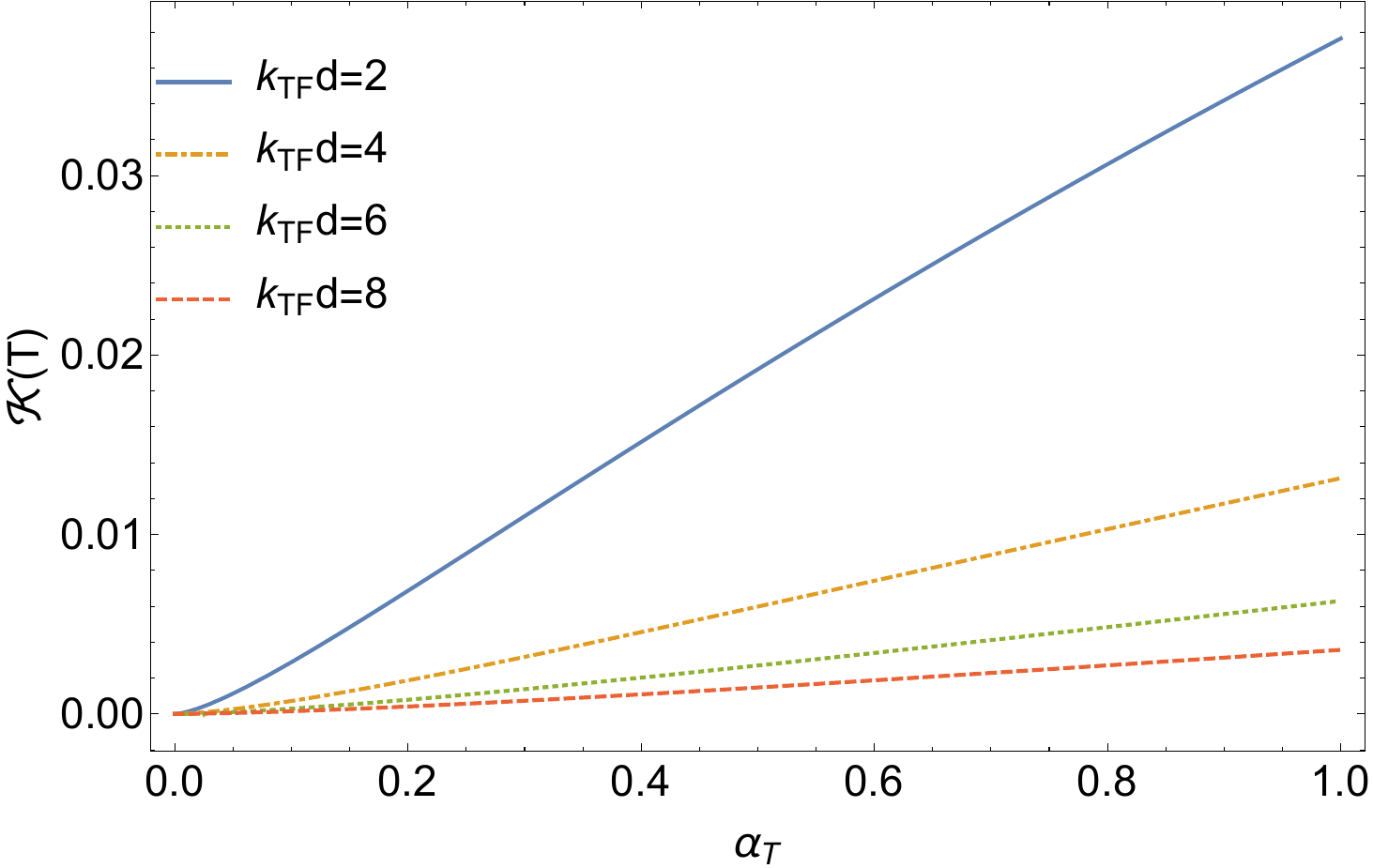}
\caption{Temperature dependence of the dimensionless thermal transfer conductance $\mathcal{K}=\varkappa/\varkappa_0$ [from Eq. \eqref{eq:kappa}] plotted in the units of $\varkappa_0=r^2_s\nu v_{\text{F}}/8\pi\phi^2k_{\text{F}}d^3$. The bottom panel shows the same data as the top panel but zoomed into the low-temperature domain of parameters.}
\label{fig:K}
\end{figure}

To this end, we turn to the analysis of Eq. \eqref{eq:kappa}. The integral in Eq. \eqref{eq:kappa}, with an input from Eqs. \eqref{eq:Pi} and \eqref{eq:Pi-U}, suggests the following dimensionless variables $x=qd$ and $y=\omega/T$ for momenta and frequencies respectively. After some straightforward algebra, this leads to the result
\begin{equation}\label{eq:K-CF}
\varkappa=\frac{Tk^2_{\text{TF}}}{8\pi^2}\mathcal{F}(\alpha_T,k_{\text{TF}}d),
\end{equation}
where we introduced the inverse Thomas-Fermi screening radius $k_{\text{TF}}=2\pi e^2\nu/\epsilon$ and dimensionless parameter $\alpha_T=2\pi\nu\phi^2Tk_Fd^3$. The two-parameter dimensionless function $\mathcal{F}(a,b)$ is defined by the following double integral 
\begin{equation}\label{eq:F}
\mathcal{F}(a,b)=\iint\limits^{\,\,\,\,\,\,+\infty}_{0}\frac{a^2x^5y^4e^{-2x}dxdy}{\sinh^2(y/2)|X_+(a,b)|^2|X_-(a,b)|^2}
\end{equation}
with
\begin{equation}
X_\pm(a,b)=x^3+bx^2(1\pm e^{-x})-iay.
\end{equation}
We have evaluated this function numerically in order to extract the characteristic dependence of thermal conductance on temperature and interlayer separation. In Fig. \ref{fig:F-cf} we plot $\mathcal{F}$ as a function of $k_{\text{TF}}d$ at several different values of $\alpha_T$, and conversely, as a function of $\alpha_T$ at varying $k_{\text{TF}}d$. We always work under the tacit assumption that $k_{\text{TF}}d>1$. This parameter range is relevant experimentally and compatible with the condition required for the near-field effect, namely $k^{-1}_{\text{TF}}<d<c/T$. It is clear that $\mathcal{F}$ decays algebraically as a power law in $1/d$, so does the NFHT conductance $\varkappa\propto (1/d)^{p_d}$ with the exponent $p_d>1$. The temperature dependence of $\mathcal{F}$, implicit in the parameter $\alpha_T$, is nonmonotonic. It has a peak at $\alpha_T\sim1$ that sets the scale for a characteristic crossover temperature $\sim E_F/(k_Fd)^3$. For $k_Fd\gg1$ this temperature is much smaller than the Fermi energy so that the crossover occurs within the domain of validity of low-energy effective model. In order to highlight the temperature dependence of conductance $\varkappa$, we plot the product $T\mathcal{F}(\alpha_T,k_{\text{TF}}d)$ versus $\alpha_T$ with the proper prefactor based on Eq. \eqref{eq:K-CF}. This is shown in Fig. \ref{fig:K}. From the graph it is apparent that thermal conductance displays monotonic growth. At lowest temperatures, it scales as a power law, $\varkappa\propto T^{p_T}$, with the exponent $p_T>1$. This regime is rather narrow. It is followed by a much wider regime with approximately $T$-linear behavior and a round off at higher temperatures when $\alpha_T>1$.        

The limiting cases discussed above can be analyzed analytically. The integral is dominated by values $\{x,y\}<1$. Provided $\{k_{\text{F}}d,k_{\text{TF}}d\}>1$ one can approximate $X_+\approx 2k_{\text{TF}}dx^2-i\alpha_Ty$ and $X_+\approx(1+ k_{\text{TF}}d)x^3-i\alpha_Ty$, and take $e^{-2x}\approx 1$ in the numerator of Eq. \eqref{eq:F}. These steps give 
\begin{align}\label{eq:F-app}
&\mathcal{F}\approx\iint\limits^{\,\,\,\,\,\,+\infty}_{0}dxdy\frac{y^4}{\sinh^2(y/2)}\nonumber \\ 
&\times \frac{\alpha^2_Tx^5}{[4(k_{\text{TF}}d)^2x^4+\alpha^2_Ty^2][(1+k_{\text{TF}}d)^2x^6+\alpha^2_Ty^2]}.
\end{align}  
For the sufficiently small $\alpha_T$, one can further neglect $\alpha^2_Ty^2$ as compared to $4(k_{\text{TF}}d)^2x^4$ in the first bracket of the denominator, but the $y$ dependence must be retained in the second bracket, which controls convergence of the $x$ integral. Note this bracket corresponds to the $\omega_-$ mode. After this additional approximation the $x$ integration becomes elementary and can be done by rescaling the variable $x\to(\alpha_Ty/(1+k_{\text{TF}}d))^{1/3}x$:
\begin{equation}
\mathcal{F}\approx\frac{1}{(2k_{\text{TF}}d)^2}\left(\frac{\alpha_T}{1+k_{\text{TF}}d}\right)^{\frac{2}{3}}\int\limits^{\infty}_{0}\frac{y^{\frac{8}{3}}dy}{\sinh^2(y/2)}
\int\limits^{\infty}_{0}\frac{xdx}{(x^6+1)}.
\end{equation}  
The integral over $x$ brings a factor of $\pi/(3\sqrt{3})$ while the $y$ integral can be expressed as a product of Euler's gamma function and Riemann's zeta function as $4\Gamma(\frac{11}{3})\zeta(\frac{8}{3})$.  Using this asymptote of $\mathcal{F}$ in Eq. \eqref{eq:kappa} we find 
\begin{equation}\label{eq:kappa-low-T}
\varkappa\approx\frac{\Gamma(\frac{11}{3})\zeta(\frac{8}{3})}{24\sqrt{3}\pi}\frac{T}{d^2}\left(\frac{\alpha_T}{1+k_{\text{TF}}d}\right)^{\frac{2}{3}}.
\end{equation} 
This asymptote applies for $T\ll E_{\text{F}}(k_{\text{TF}}d)/(k_{\text{F}}d)^3$. Recall that $k_{\text{TF}}\sim k_{\text{F}}$ for $r_s=e^2/\epsilon v_{\text{F}}\sim1$. Since $\alpha_T$ contains one power of temperature and three powers of interlayer separation  we deduce that $\varkappa\propto T^{5/3}/d^{2/3}$ in this limit. It can be compared to the corresponding result of the Fermi liquid theory $\varkappa\propto T^3/d^2$ (an extra log factor is omitted for brevity, see Refs. \cite{Kamenev-LTP,Wise-PRB21} for a complete result). We conclude that weak screening leads to a much slower decay of the heat transfer conductance with interlayer spacing thus providing a significantly stronger effect (for numerical estimates see Sec. \ref{sec:estimates}). 

Equation \eqref{eq:kappa-low-T} should also hold for the filling fractions $1/4$ and $3/4$ per layer. The only modification is in the value of the flux attachment $\phi=4$. This leads to a verifiable conclusion that at low temperatures, where Eq. \eqref{eq:kappa-low-T} applies, if the filling fraction is varied from $1/2$ to $1/4$ (or $3/4$) at fixed electron density, then the interlayer heat conductance should increase by a factor close to $4^{5/3}$.

At higher temperatures, $T>E_{\text{F}}/(k_{\text{F}}d)^2$, we deduce from Eq. \eqref{eq:F-app} a different asymptote that translates to the conductance in the form (with the logarithmic accuracy) 
\begin{equation}\label{eq:kappa-high-T}
\varkappa\approx\frac{T}{18d^2}\ln\left(\frac{\alpha_T}{k_{\text{TF}}d}\right), 
\end{equation} 
corresponding to the approximate $T$-linear regime clearly visible in Fig. \eqref{fig:K}. Interestingly, in this limit the heat conductance is nearly universal, i.e. independent of any microscopic parameters of the material (modulo the logarithmic factor).

The validity of Eq. \eqref{eq:kappa} should extrapolate to the onset of the collision-dominated regime with respect of the intralayer collisions. It is marked by the condition when the composite fermion mean free path $l_{\text{cf}}$ becomes comparable to the interlayer spacing, $l_{\text{cf}}\sim d$. In the framework of the Chern-Simons theory $l_{\text{cf}}\sim k^{-1}_{\text{F}}(E_{\text{F}}/T)^{2/3}$, therefore the approach based on the RPA breaks down above the scale of $\sim E_{\text{F}}/(k_{\text{F}}d)^{3/2}$.


\section{Disordered bilayers}\label{sec:Dis}

It is of practical importance to consider effect of disorder which is inevitably present in any bilayer device. The impurity scattering can be included via the Boltzmann equation (BE) for the composite fermion distribution function \cite{Simon-PRB93,Mirlin-PRL97}. At the simplest level, the collision term of the BE can be taken in the relaxation time approximation. While it may be insufficient in general this approach is adequate to describe the low-energy diffusive limit. Solving this equation in response to the alternating electric field $\propto e^{i\bm{qr}-i\omega t}$ leads to the density response function of the form  
\begin{equation}
\Pi^{\text{cf}}_{\rho\rho}=\frac{m^*}{2\pi}\left[1+\frac{i\omega\tau_{\text{cf}}}{\sqrt{(1-i\omega\tau_{\text{cf}})^2+(qv_{\text{F}}\tau_{\text{cf}})^2}-1}\right],
\end{equation}
where $\tau_{\text{cf}}$ is the composite fermion transport mean free time. Extrapolating this result empirically to the diamagnetic term one can write 
\begin{align}
&\Pi^{\text{cf}}_{jj}=-\frac{q^2}{24\pi m^*}\nonumber\\ 
&-\frac{i\omega m^*}{2\pi q^2\tau_{\text{cf}}}\left[(1-i\omega\tau_{\text{cf}})-\sqrt{(1-i\omega\tau_{\text{cf}})^2+(qv_{\text{F}}\tau_{\text{cf}})^2}\right],
\end{align}
which has a correct form of Eq. \eqref{eq:Pi-rho-j} in the clean limit $\tau_{\text{cf}}\to\infty$. In contrast, in the diffusive limit $\{\omega\tau_{\text{cf}},qv_{\text{F}}\tau_{\text{cf}}\}\ll1$,
expanding both $\Pi^{\text{cf}}_{\rho\rho}$ and $\Pi^{\text{cf}}_{jj}$ and using Eq. \eqref{eq:Pi-el} one finds 
\begin{equation}\label{eq:Pi-el-dif}
\Pi(q,\omega)\approx\frac{\nu Dq^2}{Dq^2-i\omega}, 
\end{equation}
with the effective diffusion constant in the form 
\begin{equation}
D=\frac{1+\phi^2/12}{1+(\phi k_{\text{F}}l_{\text{cf}}/2)^2}\frac{v^2_F\tau_{\text{cf}}}{2}. 
\end{equation}
Coincidently, the density-density response function of the form of Eq. \eqref{eq:Pi-el-dif} is identical to that of a disorder 2DEG. 
Therefore, the dynamically screened interaction in this case takes a familiar form from the theory of disordered electron systems \cite{AA}
\begin{equation}\label{eq:U-diff}
U(q,\omega)\approx\left(\frac{1}{2\nu Dq^2}\right)\frac{(Dq^2-i\omega)^2}{(1+k_{\text{TF}}d)Dq^2-i\omega}.
\end{equation}
Substituting this expression and Eq. \eqref{eq:Pi-el-dif} into Eq. \eqref{eq:kappa} leads to the heat conductance in the form 
\begin{equation}
\varkappa=\int\frac{(\omega^2/4T)^2}{\sinh^2(\omega/2T)}\frac{d\Gamma_{q\omega}}{(1+k_{\text{TF}}d)^2(Dq^2)^2+\omega^2}
\end{equation}
After integration we find as a final result 
\begin{equation}
\varkappa=\frac{3\zeta(3)}{16\pi}\frac{1}{(1+k_{\text{TF}}d)}\frac{T}{L^2_T},
\end{equation}
where we have introduced the diffusive thermal length $L_T=\sqrt{D/T}$. Notice that in this limit diffusive spreading of charge density fluctuations restores the Fermi-liquid form 
of the temperature dependence of $\varkappa\propto T^2$ \cite{Kamenev-LTP,Hekking-PRB20}. 


\section{Hydrodynamic regime}\label{sec:Hydro}

In clean electron systems with sufficiently frequent collisions the system can attain a hydrodynamic limit. It can be described macroscopically 
based on the equations of motion for particle, entropy, and momentum densities of the fluid. This formulation enables obtaining results that go beyond the perturbation theory in interaction. However, it requires a proper values of temperatures and particle density  to justify the hydrodynamic description. In the context of bilayers, hydrodynamic theory can
be applicable at temperatures when intralayer mean free path becomes shorter than interlayer separation.     

The near-field effect in the hydrodynamic limit was considered recently in Ref. \cite{AL-NFHT} for pristine systems. Here we merely repeat this calculation 
for composite fermions including the generalization to incorporate the disorder potential with the long correlation radius that exceeds the scale of equilibration length. 
It introduces friction that damps collective excitations at long wave length. In what follows, we summarize the main steps of derivation and present the end result. 
 
\begin{figure}[t!]
\centering
\includegraphics[width=3in]{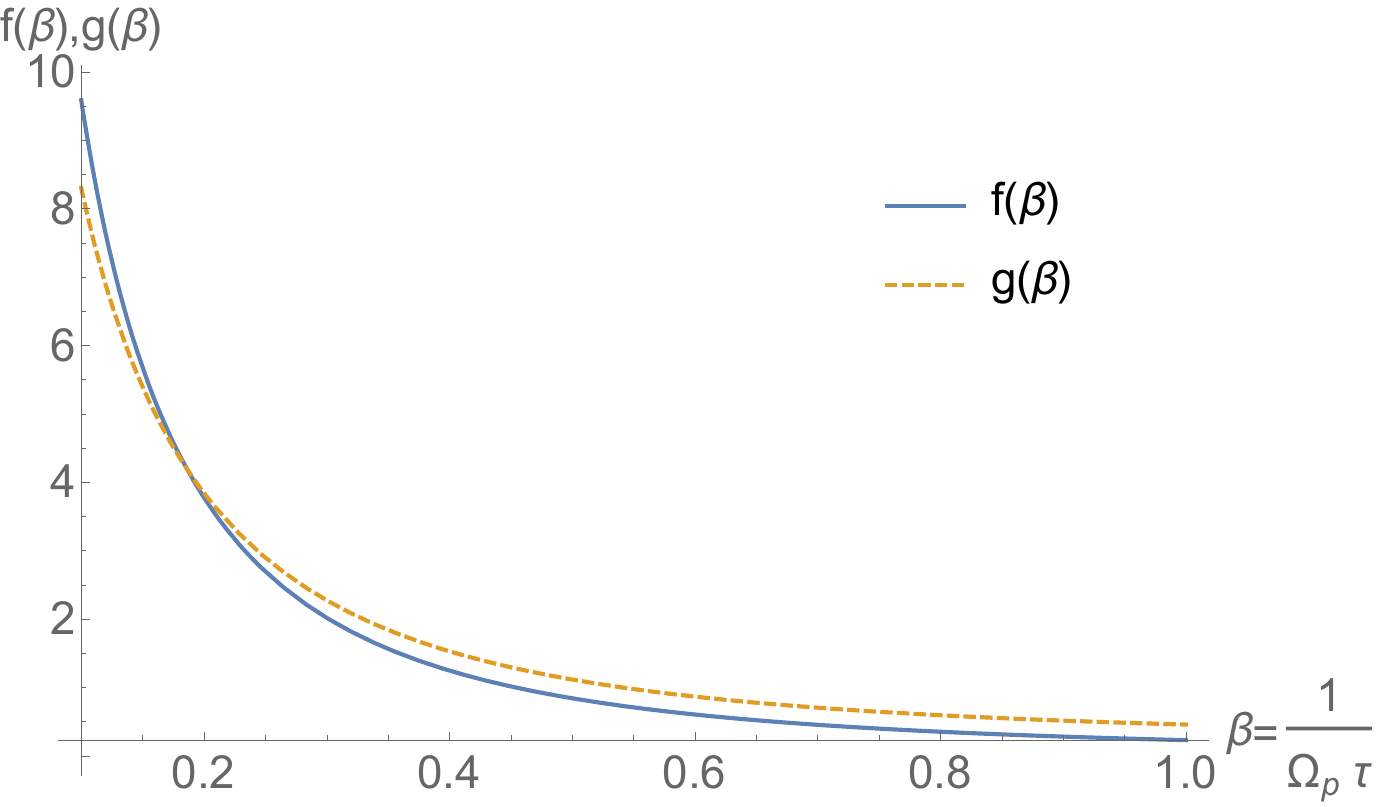}
\caption{Numerically evaluated dimensionless functions that describe the plasmon enhancement of thermal conductance $\varkappa$. 
Plasmon resonances are most pronounced for $\Omega_p\tau>1$ therefore we constrain the plot only to the values of $\beta<1$. }
\label{fig:f-g}
\end{figure}
 
One starts from the continuity and Navier-Stokes equations for an electron fluid. The latter includes electric potential, which is related to the electron charge density by the Poisson equation. Working to the linear order in fluctuations one expresses electron density variation mediated by fluctuating viscous stresses and intrinsic currents. This approach to the problem  based on the hydrodynamic equations with random Langevin fluxes is analogous to the fluctuational electrodynamics in the Coulomb limit. The correlation functions of the hydrodynamic Langevin sources are found from the respective fluctuation-dissipation relations \cite{LL-V9}. Heat flux between the layers can be directly related to the dynamic structure factor of the fluid which is given by the density-density response function \cite{Forster}. Since fluctuations of viscous stresses are not correlated with fluctuations of intrinsic currents, the corresponding near-field transfer conductance is found as a sum of two terms     
\begin{subequations}
\begin{equation}
\varkappa=\varkappa_\sigma+\varkappa_\upsilon,
\end{equation}
where 
\begin{align}\label{eq:kappa-sigma}
&\varkappa_\sigma=\int d\Gamma_{q\omega}\left(\frac{2\pi e^2}{\epsilon q}\right)e^{-qd}\left(\frac{2\sigma q^2}{e^2}\right)\nonumber \\ 
&\times\frac{\omega^2(\omega^2+1/\tau^2)(\omega^2_+-\omega^2_-)/\tau}{[(\omega^2-\omega^2_+)^2+\omega^2/\tau^2][(\omega^2-\omega^2_-)^2+\omega^2/\tau^2]},
\end{align}
and 
\begin{align}\label{eq:kappa-eta}
&\varkappa_\upsilon=\int d\Gamma_{q\omega}\left(\frac{2\pi e^2}{\epsilon q}\right)e^{-qd}\left(\frac{4(\eta+\zeta)q^4}{(m^*)^2}\right)\nonumber \\ 
&\times\frac{\omega^2(\omega^2_+-\omega^2_-)/\tau}{[(\omega^2-\omega^2_+)^2+\omega^2/\tau^2][(\omega^2-\omega^2_-)^2+\omega^2/\tau^2]}. 
\end{align}
\end{subequations}
Both contributions have transparent physical meaning. The first term under the integral of each expression, $2\pi e^2/\epsilon q$, is just the Coulomb potential that couples the layers, whereas exponential, $e^{-qd}$, captures the screening. The next factor in each term represents the strength of Langevin fluxes that drive the density fluctuations. For    
$\varkappa_\sigma$ it scales with the intrinsic conductivity of the fluid $\sigma$, whereas for $\varkappa_\upsilon$ it is determined by the shear ($\eta$) and bulk ($\zeta$) viscosities, as dictated by the fluctuation-dissipation theorem. For the case of short-ranged interactions of composite fermions one can extract intrinsic conductivity and viscosity from the   
model of Fermi surface coupled to $U(1)$ gauge field for dynamical critical exponent $z=3$. The dissipative coefficients have the following parametric dependence \cite{HLR,Davison:2015,Eberlein-PRB17}:
\begin{equation}
\sigma\propto e^2\left(\frac{E_{\text{F}}}{T}\right)^{2/3},\quad \eta\sim\zeta\propto k^2_{\text{F}}\left(\frac{E_{\text{F}}}{T}\right)^{2/3}.
\end{equation} 
The remaining terms in Eqs. \eqref{eq:kappa-sigma} and \eqref{eq:kappa-eta} are just the corresponding parts of the dynamical structure factor that are peaked at the frequencies of two plasmon modes
\begin{equation}
\omega^2_\pm=\frac{2\pi ne^2q}{\epsilon m^*}(1\pm e^{-qd}).
\end{equation}  
The broadening of these resonances is governed by the relaxation time $\tau$ induced by the disorder potential. The primary difference with the earlier analysis is that in pristine systems decay of plasmons is determined by the combination of viscous diffusion, $\Im \omega\propto (\eta+\zeta)q^2/(m^*n)$, and the Maxwell relaxation, $\Im\omega\propto \sigma q$. It is clear that in the long wave length limit, $q\to0$, attenuation of plasmons is ultimately determined by disorder potential \cite{Zverevich-Plasmons}.   

Both expressions for $\varkappa_\sigma$ and $\varkappa_\upsilon$ can be significantly simplified by rescaling momentum integrals in units of $x=qd$ and frequency integrals in units of $y=\omega\tau$. This gives as a result 
\begin{equation}\label{eq:kappa-hydro}
\varkappa_\sigma=\frac{\sigma}{\epsilon d^3}f(\beta),\quad 
\varkappa_\upsilon=\frac{\upsilon}{d^4}g(\beta).
\end{equation}
where we introduced the kinematic viscosity of the fluid $\upsilon=(\eta+\zeta)/nm^*$. The dimensionless functions $f$ and $g$ depend on a single variable $\beta=1/(\Omega_p\tau)$, where $\Omega_p=\sqrt{2\pi e^2n/\epsilon m^*d}$ is the plasmon frequency at the characteristic momentum transfer $q=1/d$. For plasmons to be well-defined and long-lived excitation we assume $\Omega_p\tau>1$. These functions capture the plasmon enhancement of the heat transfer. They can be evaluated numerically and plotted in Fig. \ref{fig:f-g}. A crude estimate suggests $f\sim g\sim \ln^4(1/\beta)$ for $\beta\ll1$. Finally, we should note that the crossover from the low-temperature RPA-limit of the heat transfer given by Eq. \eqref{eq:kappa-high-T} to the higher-temperature hydrodynamic limit given by Eq. \eqref{eq:kappa-hydro} is not immediately clear and requires additional consideration.


\begin{figure}[t!]
\centering
\includegraphics[width=3in]{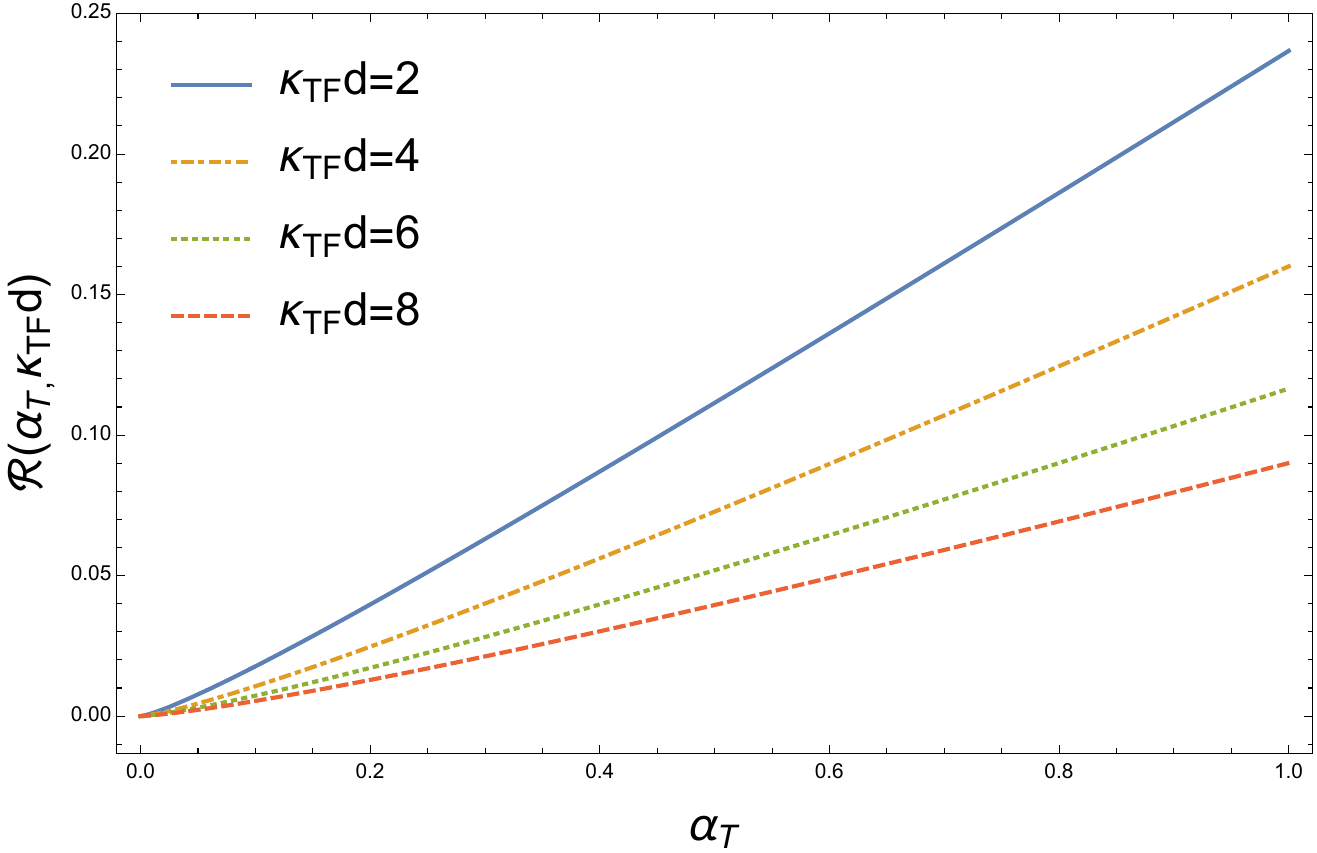}
\caption{Temperature dependence of the drag resistance presented in the dimensionless units. Based on Eq. \eqref{eq:R} $\mathcal{R}$ is defined by $\rho_D$ in units of $(1/4\pi e)^2(1/nd^2)^2$ whereas $\alpha_T\propto T$. The domain of validity of Eq. \eqref{eq:rhoD-cf-1}, when $\mathcal{R}\propto T^{4/3}$, is limited to the lowest temperatures $T\ll E_{\text{F}}/(k_{\text{F}}d)^2$. Above this scale drag resistance is nearly linear in temperature $\mathcal{R}\propto T\ln T$, see Eq. \eqref{eq:rhoD-cf-2}.}
\label{fig:R}
\end{figure}

\section{Drag resistance}\label{sec:Drag}

This section serves a complementary purpose. First, we recapitulate known result for the drag resistance in electron bilayers at half filling of the Landau level \cite{Ussishkin-PRB97,Sakhi-PRB97,Ussishkin-PRL98,Millis-PE99}. Next, we highlight several experimentally relevant limiting cases which were not discussed in the literature before. We do so in light of the near heat field transfer problem, which originates from the same mechanism of interlayer coupling. Therefore we relate the two phenomena in the context of composite fermions. 
This parallels recent discussion of the related physics presented in the context of strange metals \cite{Chudnovskiy}.  

Coulomb drag \cite{CD-Review} is an example of a nonlocal transport effect that arises in the electrically isolated but interactively coupled conducting layers when drag voltage $V_{\text{drag}}$ is induced in one layer by the drive current $I_{\text{drive}}$ in the other layer. The nonlocal resistance $\rho_D=V_{\text{drag}}/I_{\text{drive}}$ provides a direct measure of the interlayer electron correlations. The microscopic foundations of the drag effect were developed in Refs. \cite{JauhoSmith,MacDonald,KamenevOreg,Flensberg-PRB95} and the corresponding resistance can be expressed as follows 
\begin{equation}\label{eq:rhoD}
\rho_D=\frac{1}{2e^2n^2T}\int \frac{q^2(\Im\Pi(q,\omega))^2|U(q,\omega)|^2}{\sinh^{2}(\omega/2T)}d\Gamma_{q\omega}. 
\end{equation}
Apart from the overall factor, $\rho_D$ has almost the same form as the expression for $\varkappa$ in Eq. \eqref{eq:kappa} except that under the integral two powers of frequency, relevant for the heat transfer, are replaced by the two powers of momentum, relevant for the drag. It is therefore not surprising that these two quantities share similar characteristic features.    

\subsection{Ballistic limit}

We use Eqs. \eqref{eq:Pi} and \eqref{eq:Pi-U} to rewrite Eq. \eqref{eq:rhoD} in the dimensionless notations 
\begin{equation}
\rho_D=\frac{1}{8\pi^2 e^2}\frac{1}{(nd^2)^2}\mathcal{R}(\alpha_T,k_{\text{TF}}d)
\end{equation}
where 
\begin{equation}\label{eq:R}
\mathcal{R}(a,b)=\iint\limits^{\,\,\,\,\,\,+\infty}_{0}\frac{a^2b^2x^7y^2e^{-2x}dxdy}{\sinh^2(y/2)|X_+(a,b)|^2|X_-(a,b)|^2}
\end{equation}
with $x=qd$ and $y=\omega/T$. From here one can extract two limiting case of interest. At lowest temperatures, $T\ll E_{\text{F}}/(k_{\text{F}}d)^2$, one can use exactly the same approximations that lead to Eq. \eqref{eq:F-app}, namely, take $X_+\approx 2k_{\text{TF}}dx^2$, $X_-\approx(1+k_{\text{TF}}d)x^3-i\alpha_Ty$, and $e^{-2x}\approx 1$, which gives
\begin{equation}
\rho_D\approx\frac{\alpha^2_T}{32\pi^2e^2}\frac{1}{(nd^2)^2}\iint\limits^{\,\,\,\,\,\,+\infty}_{0}\frac{x^3y^2dxdy}{\sinh^2\frac{y}{2}[(1+k_{\text{TF}}d)^2x^6+\alpha^2_Ty^2]}
\end{equation}
Here $x$ integral can be done first by rescaling the variable, followed by $y$ integral. Collecting all the factors we get 
\begin{equation}\label{eq:rhoD-cf-1}
\rho_D\approx\frac{\Gamma(\frac{7}{3})\zeta(\frac{4}{3})}{24\sqrt{3}\pi e^2}\left(\frac{1}{nd^2}\right)^2\left(\frac{\alpha_T}{k_{\text{TF}}d}\right)^{\frac{4}{3}}
\end{equation}
which reproduces earlier result $\rho_D\propto T^{4/3}$ \cite{Ussishkin-PRB97,Sakhi-PRB97,Ussishkin-PRL98,Millis-PE99}. At the intermediate temperatures, when $\alpha_T>k_{\text{TF}}d$, we use different approximations to extract the leading asymptote.  The expressions of $X_{\pm}$ remain the same, but we recognize that the dominant range of $y$ integration is limited by the domain $y\ll1$ therefore we can take $y^2/\sinh^2\frac{y}{2}\to4$. This gives 
\begin{align}
&\rho_D\approx\frac{\alpha^2_T}{2\pi^2e^2}\frac{(k_{\text{TF}}d)^2}{(nd^2)^2}\nonumber \\ 
&\times \iint\limits^{\,\,\,\,\,\,+\infty}_{0}\frac{x^7e^{-2x}dxdy}{[4(k_{\text{TF}}d)^2x^4+\alpha^2_Ty^2][(1+k_{\text{TF}}d)^2x^6+\alpha^2_Ty^2]}.
\end{align}
At this point $y$ integration can be completed exactly with the help of the tabulated integral for the product of two Loretzians
\begin{equation}
\int^{\infty}_{0}\frac{dy}{(y^2+a^2)(y^2+b^2)}=\frac{\pi}{2ab(a+b)}, 
\end{equation}  
and the remaining $x$ integral can be done with the logarithmic accuracy to yield the final expression 
\begin{equation}\label{eq:rhoD-cf-2}
\rho_D\approx\frac{1}{12\pi  e^2}\left(\frac{1}{nd^2}\right)^2\left(\frac{\alpha_T}{k_{\text{TF}}d}\right)\ln\left(\frac{\alpha_T}{k_{\text{TF}}d}\right).
\end{equation}
In this limit we deduce $\rho_D\propto T\ln T$. This behavior was clearly observed in several experiments \cite{Lilly-PRL98,Klitzing-PE00,Tutuc-PRB09}, 
however it was not addressed theoretically. The full temperature dependence is depicted in Fig. \ref{fig:R} for several values of $k_{\text{TF}}d$.  

\subsection{Diffusive limit}

Eq. \eqref{eq:rhoD} for the drag resistance applies to the disordered systems as well with the proper modifications to the polarization function and screened interaction. 
Therefore, using Eqs. \eqref{eq:Pi-el-dif} and \eqref{eq:U-diff} we find 
\begin{equation}
\rho_D=\frac{1}{8e^2n^2T}\int\frac{\omega^2}{\sinh^2\frac{\omega}{2T}}\frac{q^2d\Gamma_{q\omega}}{(1+k_{\text{TF}}d)^2(Dq^2)^2+\omega^2}.
\end{equation}
The resulting temperature dependence can be estimated as follows. It is clear that the leading contribution comes the
low-frequency $\omega\lesssim T$ and small-wave-vector behavior of the integrand. 
Thus the contributions to the integral in $q$ for $q<\sqrt{\omega/D}$ can be neglected. These observations lead to the approximate expression 
\begin{equation}
\rho_D\approx \frac{1}{32\pi^2e^2n^2}\frac{1}{(1+k_{\text{TF}}d)^2}
\int\frac{\omega^2d\omega/T}{\sinh^2\frac{\omega}{2T}}\int\limits^{1/l_{\text{cf}}}_{\sqrt{\omega/D}}\frac{dq}{D^2q}.
\end{equation}
The logarithmically divergent momentum integral is cut by the inverse of the mean free path at the upper limit, which confines the applicability of the 
diffusive approximation, and by thermal length at the lower limit, due to approximation made to the screening of the interaction.
In principle, frequency integral should also be stopped at $1/\tau_{\text{cf}}$, but owing to its rapid convergence the limits can be extended to infinity.  
Therefore, with the logarithmic accuracy, one arrives at  
\begin{equation}
\rho_D\approx\frac{1}{12e^2}\left(\frac{1}{nL^2_T}\right)^2\left(\frac{1}{1+k_{\text{TF}}d}\right)^2\ln\frac{L_T}{l_{\text{cf}}}.
\end{equation}
Coincidently, the resulting temperature dependence is identical to that of drag resistance derived for the disordered Fermi gas at zero field $\rho_D\propto T^2\ln T$ \cite{MacDonald,KamenevOreg}. 

\subsection{Hydrodynamic limit}

Drag effect also admits hydrodynamic description in strongly correlated electron liquids \cite{Patel-PRB17,Apostolov-PRB14,Holder:2019,Stoof:2022}. Computation of the dragging force exerted from the drive layer on electron fluid in the drag layer requires considerations of the density fluctuation advected by the flow. This analysis can be carried out to the linear order in hydrodynamic velocity. We find that drag resistance is dominated by plasmon resonance and can be expressed as the sum of two terms 
\begin{equation}
\rho_D=\rho_\sigma+\rho_\upsilon.
\end{equation}
In complete analogy with the heat transfer conductance, the first term $\rho_\sigma$ is generated by fluctuating intrinsic currents, whereas the second term $\rho_\upsilon$ stems from the density fluctuations induced by viscous stresses. We find that the latter is parametrically smaller and decays fast with the interlayer separation. Therefore, we focus on the former contribution that can be found in the following analytical form    
\begin{align}\label{eq:rhoD-sigma}
&\rho_\sigma=\frac{1}{2e^2n^2}\int d\Gamma_{q\omega}\left(\frac{2\pi e^2}{\epsilon q}\right)e^{-qd}\left(\frac{T\sigma q^4}{e^2}\right)\nonumber \\ 
&\times\frac{\omega^2(\omega^2_+-\omega^2_-)/\tau}{[(\omega^2-\omega^2_+)^2+\omega^2/\tau^2][(\omega^2-\omega^2_-)^2+\omega^2/\tau^2]}.
\end{align}
Observe that the splitting of plasmon resonances becomes exponentially small at $q>1/d$. Therefore, in order to estimate the frequency integral in the above expression it is not sufficient to take poles of two separate Lorentzians. Fortunately, this integral can be calculated exactly so that the above expression reduces to 
\begin{align}
&\rho_\sigma=\frac{1}{2e^2n^2}\int \frac{d^2q}{(2\pi)^2}\left(\frac{2\pi e^2}{\epsilon q}\right)e^{-qd}\left(\frac{T\sigma q^4}{e^2}\right)\nonumber \\ 
&\times\frac{(\omega^2_+-\omega^2_-)}{[(\omega^2_+-\omega^2_-)^2+2(\omega^2_++\omega^2_-)/\tau^2]}.
\end{align}
The final result for the transresistance can be presented in the form   
\begin{equation}
\rho_\sigma=\frac{\sigma}{e^4}\frac{T}{E_{\text{F}}}\frac{1}{(nd^2)^2}h(\beta),\quad \beta=\frac{1}{\Omega_p\tau}.
\end{equation}
The dimensionless function is defined by the integral 
\begin{equation}
h(\beta)=\frac{1}{4}\int^{\infty}_{0}\frac{x^4e^{-2x}dx}{xe^{-2x}+1/\beta^2}.
\end{equation} 
This contribution to drag resistance displays anomalous sublinear temperature dependence, $\rho_D\propto T^{1/3}$. This behavior is qualitatively consistent with experimental observations reported in Refs. \cite{Tutuc-PRB09,Klitzing-PE00}. Finally, we deduce that viscosity dependent contribution scales as follows
\begin{equation}
\rho_\upsilon=\frac{\epsilon v_{\text{F}}}{e^4}\frac{T}{E_{\text{F}}}\frac{\eta+\zeta}{n}\frac{1}{(k_{\text{F}}d)^5}w(\beta),
\end{equation}
where $w(\beta)$ is yet another dimensionless function that has logarithmic dependance on $\beta$ in the well-resolved plasmon limit when $\beta<1$.


\section{Parameters and estimates}\label{sec:estimates}

Drag of composite fermions was measured in several groundbreaking experiments \cite{Lilly-PRL98,Gramila-PRL00,Klitzing-PE00,Tutuc-PRB09}. We take some typical values for parameters of the bilayer devices to estimate the magnitude of the effect and compare that to drag between weakly correlated 2D electron systems. Near field effect was not measured for composite fermion bilayers but we will assume the same range of parameters in order to estimate its value comparatively to the known examples. Assuming identical layers with the average electron density $n\sim 10^{11}$ cm$^{-2}$ and interlayer spacing $d\sim 200 \AA$, we can estimate $k_{\text{F}}d\sim 2$. In principle, this product can be much bigger since drag resistance was successfully measured for much larger interlayer separations up to $d\sim5000 \AA$ \cite{Gramila-PRB93}. For the effective mass of composite fermions we take $m^*\approx 12m_b$, where $m_b\approx0.067m_e$ is the band mass in GaAs and $m_e$ is bare electron mass. Putting these numbers together we estimate the Fermi energy to be about $E_{\text{F}}\sim 10$ K. The condition $\alpha_{T^*}=k_{\text{TF}}d$ defines the crossover temperature $T^*$ between low and intermediate temperature regimes. It evaluates to $T^*=(2e^2/\phi^2\epsilon v_{\text{F}})E_{\text{F}}/(k_{\text{F}}d)^2$  and for above parameters is close to $\sim 1$ K. In a weakly correlated 2DEG bilayer drag resistance is given by $\rho_D\sim e^{-2}(T/E_{\text{F}})^2/(k_{\text{F}}d)^4$ at temperatures $T<E_{\text{F}}/(k_{\text{F}}d)$. If we compare that to resistance of composite fermions at $T\sim T^*$ from Eq. \eqref{eq:rhoD-cf-2}, accounting for the difference in Fermi energy due to effective mass, we find the ratio of the two to be roughly $(m^*/m_b)^2(k_{\text{F}}d)^4\sim 10^3$. 
Therefore, drag of composite fermions is three orders of magnitude stronger. This analysis corroborates earlier conclusions \cite{Ussishkin-PRB97}.
The composite fermion drag measured in approximately equivalent bilayers (in terms of values of $n$ and $d$) give values $\rho_D\sim 100\Omega/\square$ \cite{Lilly-PRL98} and $\rho_D\sim 2\Omega/\square$ \cite{Klitzing-PE00} at $T\sim 2$K. The typical number of the zero-field value of drag in weakly correlated 2DEG bilayers is in the range of $\rho_D\sim 10$ m$\Omega/\square$ \cite{Gramila-PRL91} at $T\sim2$ K. Thus, strong enhancement of drag in composite fermions systems is apparent.  

Interestingly, the same enhancement parameter applies to the near field thermal conductance. Indeed, the Fermi liquid prediction $\varkappa\sim (T/d^2)(T/E_{\text{F}})^2$ applicable for $T<E_{\text{F}}/(k_{\text{F}}d)$ should be compared to Eq. \eqref{eq:kappa-low-T} or \eqref{eq:kappa-high-T} depending on the value of $T$. Then for $T\sim T^*$ we deduce the same parametric enhancement given by $(m^*/m_b)^2(k_{\text{F}}d)^4\gg1$.    

In summary, we conducted a comprehensive analysis of the near-field thermal transfer conductance in bilayers of composite fermions. Our theoretical framework spans various transport regimes, encompassing the ballistic to hydrodynamic limits. The impact of disorder on the conductance is also discussed. Additionally, we delved into the issue of drag resistance, reproducing prior findings and expanding the scope of results to a wider range of parameters pertinent to experimental devices and measurements.


\section*{Acknowledgments}
We thank Ady Stern for the discussions of the results. This work was supported by the National Science Foundation Grant No. DMR-2203411 and H. I. Romnes Faculty Fellowship provided by the University of Wisconsin–Madison Office of the Vice Chancellor for Research and Graduate Education with funding from the Wisconsin Alumni Research Foundation. A.L. acknowledges the hospitality of Thouless Institute for Quantum Matter during the 2024 Winter Workshop on the topic of ``New Developments in Fractionalization" and the support during the visit.

\bibliography{biblio}

\end{document}